\documentclass[10pt]{JHEP3}

\usepackage{epsfig,multicol,bbm}
\usepackage{amsmath}
\usepackage{graphicx}
\usepackage{dcolumn}
\usepackage{bm}
\usepackage{amssymb}
\usepackage{latexsym}
\usepackage{multirow}

\newcommand\fverb{\setbox\fverbbox=\hbox\bgroup\verb}
\newcommand\fverbdo{\egroup\medskip\noindent%
            \fbox{\unhbox\fverbbox}\ }
\newcommand\fverbit{\egroup\item[\fbox{\unhbox\fverbbox}]}
\newbox\fverbbox

\title{Cardassian Universe Constrained by Latest Observations }

\author{Chao-Jun Feng\\
    Shanghai United Center for Astrophysics(SUCA), Shanghai Normal University,\\
    100 Guilin Road, Shanghai 200234, P.R.China\\
    E-mail: \email{fengcj@shnu.edu.cn}}

\author{Xin-Zhou Li\\
    Shanghai United Center for Astrophysics(SUCA), Shanghai Normal University,\\
    100 Guilin Road, Shanghai 200234, P.R.China\\
    E-mail: \email{kychz@shnu.edu.cn}}

\received{\today}       
\accepted{\today}       

\abstract{Several Cardassian universe models including the original, modified polytropic and exponential Cardassian models are constrained by the latest Constitution Type Ia supernova data, the position of the first acoustic peak of CMB from the five years WMAP data and the size of baryonic acoustic oscillation peak from the SDSS data. Both the spatial flat and curved universe are studied, and we also take account of the possible bulk viscosity of the matter fluid in the flat universe case.}

\keywords{Cosmology, Viscosity, Cardassian universe}

\begin{document}


\section{Introduction}
Nowadays, there are many dark energy models and modified gravity theories proposed to explain the current accelerating
expansion of the universe, which has been confirmed by the observations like Type Ia supernovae (SNe Ia), CMB and SDSS
et al. The dark energy models assume the existence of an energy component with negative pressure in the universe, and
it dominates and accelerates the universe at late times. The cosmological constant seems the best candidate of dark
energy, but it suffers the fine tuning problem and coincidence problem, and it may even have the age problem
\cite{Feng:2009jr}. To alleviate these problems, many dynamic dark energy models were proposed. However, people still
do not know what is dark energy.

Since the Einstein general gravity theory has not been checked in a very large scale, then one does not know whether
this gravity theory  is suitable or not for studying the observational data like SNe Ia, and maybe the accelerating
expansion of universe is due to the gravity theory that differs from the general gravity. Thus, many modified gravity
theories like $f(R)$, DGP et al. are proposed to explain the accelerating phenomenology. The Cardassian model is a kind
of model in which the Fridemann equation is modified by the introduction of an additional nonlinear term of energy
density, and we will briefly review on this model in the next section.

Dissipative processes in the universe including bulk viscosity, shear viscosity and heat transport have been
conscientiously studied\cite{barrow}. The general theory of dissipation in relativistic imperfect fluid was put on a
firm foundation by Eckart\cite{eckart}, and, in a somewhat different formulation, by Landau and Lifshitz\cite{landau}.
This is only the first order deviation from equilibrium and may has a causality problem, the full causal theory was
developed by Isreal and Stewart\cite{israel}, and has also been studied in the evolution of the early
universe\cite{harko}. However, the character of the evolution equation is very complicated in the full causal theory.
Fortunately, once the phenomena are quasi-stationary, namely slowly varying on space and time scale characterized by
the mean free path and the mean collision time of the fluid particles, the conventional theory is still valid. In the
case of isotropic and homogeneous universe, the dissipative process can be modeled as a bulk viscosity $\zeta$ within a
thermodynamical approach, while the shear viscosity $\eta$ can be neglected, which is consistent with the usual
practice\cite{brevik}. For works on viscous dark energy models, see ref.\cite{vde}.

The bulk viscosity introduces dissipation by only redefining the effective pressure, $p_{eff}$, according to
$p_{eff}=p-3\xi H$ where $\xi$ is the bulk viscosity coefficient and $H$ is the Hubble parameter. The condition $\xi>0$
guaranties a positive entropy production, consequently, no violation of the second law of the
thermodynamics\cite{zimdahl}. The case $\xi =\tau H$, implying the bulk viscosity is proportional to the fluid's
velocity vector, is physically natural, and has been considered earlier in  a astrophysical context, see the review
article of Gr{\o}n\cite{gron}.

In this paper, we will focus on several Cardassian models including the original, modified polytropic and exponential
Cardassian models and constrain their parameters by the latest Constitution Type Ia supernova data (SNeIa), the
position of the first acoustic peak of the cosmic microwave background (CMB) from the five years WMAP data and the size
of baryonic acoustic oscillation (BAO) peak from the SDSS data. We have consider the case of spatial flat and curved
universe and the case of flat universe with the bulk viscosity. After a lengthy numerical calculation, we obtain the
best fit values of the parameters in each Cardassian model.

This paper is organized as follows: In Section 2, we present a brief review of Cardassian models, and derive the Hubble
parameter in terms of the redshift and some parameters for several models. In Section 3, we analysis each model with
statistical method and constrain their parameters with the observational data. In the last section, we give some
conclusions and discussions.

\section{The Cardassian model}

Assuming the universe is homogeneous and isotropic, i.e.
\begin{equation}\label{metric}
    ds^2 = -dt^2 + a(t)^2 \left(\frac{dr^2}{1-kr^2} + r^2(d\theta^2 + \sin^2\theta d\phi^2)\right)\,,
\end{equation}
where $k$ is the spatial curvature, the modified Friedmann equation for Cardassian model  is given by
\begin{equation}\label{Friedmann equ}
    H^2+ \frac{k}{a^2}  = \frac{g(\rho)}{3}\,,
\end{equation}
where $\rho$ is the total energy density of matter and radiation and we will neglect the contribution of radiation for
the late-time evolution of the universe. In eq.(\ref{Friedmann equ}), the function $g(\rho_m)$  reduces to $\rho_m$ in
the early universe, then eq.(\ref{Friedmann equ}) reduces to the ordinary Friedmann equation during early epochs such
as primordial nucleosynthesis. However, it differs from the FRW universe at the redshift $z<\mathcal{O}(1)$, during
which it will gives rise to accelerated expansion. Different forms of the function $g(\rho_m)$ corresponds to different
Cardassian models, and we will focus on the original Cardassian model (OC) \cite{OC},  the modified polytropic Cardassian model (MPC) \cite{MPC}, the exponential model (EC) \cite{EC}, their flat versions ( FOC, FMPC, FEC ), in which the spatial curvature is neglected, and their viscous versions ( VOC, VMPC, VEC ) \cite{vcard}, in which the bulk viscosity of the matter is taken account and the spatial curvature is also neglected. We summaries these models in Table \ref{table:models}. Recent works on constraining the Cardassian universe, see ref. \cite{recent}.

\begin{table}[h]
\centering
  \begin{tabular}{l|c|l}
  \hline
  \hline
  \multicolumn{1}{c|}{$g(\rho_m)$} &Model & \multicolumn{1}{c}{$E^2=H^2/H_0^2$} \\
  \hline
  \hline
  \multirow{3}*{ $\rho_m\left[1+\left(\frac{\rho_m}{\rho_{card}}\right)^{n-1}\right] $  }
    & FOC &
    $\Omega_{m0}(1+z)^3 + (1-\Omega_{m0})(1+z)^{3n}$ \\
  \cline{2-3}
    & OC &
    $\Omega_{m0}(1+z)^3 + (1-\Omega_{m0}-\Omega_{k0})(1+z)^{3n}+\Omega_{k0}(1+z)^2$ \\
  \cline{2-3}
    & VOC &
    $\Omega_{m0}(1+z)^{3 (1-\tau) } \big[\frac{\tilde\tau+1}{\tilde\tau+F_{voc}(z)}\big]^{\tau/(\tau-n)}
     \left[1+F_{voc}(z) (\Omega_{m0}^{-1}-1)\right]$ \\
  \hline
  \multirow{3}*{ $\rho_m\left[1+\left(\frac{\rho_m}{\rho_{card}}\right)^{q(n-1)}\right]^{\frac{1}{q}} $}
    & FMPC &
    $\Omega_{m0}(1+z)^3[1+(\Omega_{m0}^{-q}-1)(1+z)^{3q(n-1)}]^{1/q}$\\
  \cline{2-3}
    & MPC &
    $\Omega_{m0}(1+z)^3[1+((1-\Omega_{k0})^q\Omega_{m0}^{-q}-1)(1+z)^{3q(n-1)}]^{1/q}+\Omega_{k0}(1+z)^2 $\\
  \cline{2-3}
    & VMPC &
    $\Omega_{m0}(1+z)^{3 (1-\tau) } \big[\frac{\tilde\tau_2+1}{\tilde\tau_2+F_{vmpc}(z)}\big]^{\frac{\tau}{q(\tau-n)}}
    \left[1+F_{vmpc}(z) (\Omega_{m0}^{-q}-1)\right]^{\frac{1}{q}} $\\
  \hline
  \multirow{3}*{ $\rho_m \exp{\left[\left(\frac{\rho_m}{\rho_{card}}\right)^{-n}\right]}$~ }
    & FEC &
    $\Omega_{m0}(1+z)^3\exp{[-(1+z)^{-3n}\ln\Omega_{m0}]} $\\
  \cline{2-3}
    & EC &
    $\Omega_{m0}(1+z)^3\exp{[-(1+z)^{-3n}(\ln\Omega_{m0}-\ln(1-\Omega_{k0}))]} + \Omega_{k0}(1+z)^2 $\\
  \cline{2-3}
    & VEC &
    $ \Omega_{m0}(1+z)^{3 }
    \big[\frac{\tilde\tau_3+1}{\tilde\tau_3+F_{vec}(z)}\big]^{\tau/(n(1-\tau))}
    \exp{(-F_{vec}^{-1}~\ln\Omega_{m0})}$\\
  \hline
  \end{tabular}
  \caption{\label{table:models} Summary of Cardassian models with different functions of $g(\rho_m)$. Here, $\rho_{card}$ is a character energy density. }
\end{table}

Energy conservation of pressureless matter is given by
\begin{equation}\label{law matter}
    \dot\rho_m + 3H(\rho_m - 3\xi_m H) = 0 \,,
\end{equation}
where $\xi_m$ is the bulk viscosity for the matter $\rho_m$. Following \cite{tot den}, the function $g$ could be written as $g = \rho_m + \rho_k$, where $\rho_k$ is so called Cardassian term, which may indicate that our observable universe as $3+1$ dimensional brane in extra dimensions. Thus, the total energy density can be written as
\begin{equation}\label{law tot}
    \dot g + 3H(g + p_T - 3\xi H) = 0 \,,
\end{equation}
where $\xi$ is the  bulk viscosity for the total energy density $g(\rho_m)$. Here, $p_T$ is defined as the effective pressure of total fluid without bulk viscosity, and the first law of thermodynamics in an adiabatic expanding universe gives
\begin{equation}
    p_T = \rho_m  \frac{\partial g}{\partial \rho_m} - g \,.
\end{equation}
Therefore, one can get $\xi = \frac{\partial g}{\partial \rho_m}\xi_m$ from eqs. (\ref{law matter}) and (\ref{law
tot}). In the following, we will choose $\xi = \tau H$,  in which the cosmological dynamics can be analytically
solvable \cite{vcard} and $\tau$ is a constant. Then, the conservation law (\ref{law tot}) becomes
\begin{equation}\label{law tot2}
    \left(f+ \frac{\partial f}{\partial y }\right) y' + 3 \left[ \frac{\partial f}{\partial y } + (1- \tau)f \right] = 0 \,,
\end{equation}
where $f = g/\rho_m$, $y = \ln(\rho_m/\rho_{card})$ the prime denotes the derivative with respect to $x \equiv \ln (a)
= -\ln(1+z)$, and $z$ is the redshift.

For the VOC model, $f = 1+e^{(n-1)y}$, and the solution is
\begin{equation}\label{voc sol}
     \rho_m = \rho_{m0} (1+z)^{3 (1-\tau) }
    \bigg[\frac{1-\tau+(n-\tau)\left(\rho_{m0}/\rho_{card}\right)^{n-1}
    }{1-\tau+(n-\tau)\left(\rho_m/\rho_{card}\right)^{n-1}}\bigg]^{\frac{\tau}{\tau-n}}
     \,,
\end{equation}
where $\rho_{m0}$ is the present value of the matter's energy density, and the Hubble parameter $E^2 = H^2/H_0^2$, is
given by
\begin{equation}\label{voc:hubble}
     E^2 = \Omega_{m0}(1+z)^{3 (1-\tau) }
    \bigg[\frac{\tilde\tau_1+1}{\tilde\tau_1+F_{voc}(z)}\bigg]^{\frac{\tau}{\tau-n}}
    \left[1+F_{voc}(z) (\Omega_{m0}^{-1}-1)\right]
\end{equation}
where $\Omega_{m0} = \rho_{m0}/(3H^2_0)$ , $H_0$ is the present value of the Hubble parameter and
\begin{equation}
    \tilde\tau_1 = \left(\frac{\Omega_{m0}}{1-\Omega_{m0}}\right)\left( \frac{1-\tau}{n-\tau} \right)\,.
\end{equation}
Here the function $F_{voc}(z) = (\rho_m/\rho_{m0})^{n-1}$ satisfies
\begin{equation}
    F_{voc} = (1+z)^{3 (1-\tau)(n-1) }
    \bigg[\frac{\tilde\tau_1+1 }{\tilde\tau_1+F_{voc}}\bigg]^{\frac{\tau(n-1)}{\tau-n}} \,,
\end{equation}
from which one can get the solution for $F_{voc}$ and substitute it into eq.(\ref{voc:hubble}), then one obtains the Hubble
parameter in terms of $z$ and parameters $\Omega_{m0}, n, \tau$. When $\tau = 0$, the solution is rather simple, and
the Hubble parameter (\ref{voc:hubble}) becomes
\begin{equation}\label{foc:Hubble}
    E^2 = \Omega_{m0}(1+z)^{3 } + (1-\Omega_{m0})(1+z)^{3n} \,.
\end{equation}

For the VMPC model, $f = ( 1+e^{q(n-1)y})^{1/q}$, and the solution is
\begin{equation}\label{VMPC sol}
    \rho_m = \rho_{m0} (1+z)^{3  (1-\tau )}
    \left[\frac{1-\tau + (n-\tau )(\rho_{m0}/\rho_{card})^{q(n-1)} }{1-\tau + (n-\tau )(\rho_{m}/\rho_{card})^{q(n-1)}}\right]^{\frac{\tau}{ q(\tau
    -n)}}\,,
\end{equation}
and the Hubble parameter is given by
\begin{equation}\label{vmpc:hubble}
    E^2 = \Omega_{m0}(1+z)^{3 (1-\tau) }
    \bigg[\frac{\tilde\tau_2+1}{\tilde\tau_2+F_{vmpc}(z)}\bigg]^{\frac{\tau}{q(\tau-n)}}
    \left[1+F_{vmpc}(z) (\Omega_{m0}^{-q}-1)\right]^{\frac{1}{q}} \,,
\end{equation}
where
\begin{equation}
    \tilde\tau_2 = \left(\frac{\Omega_{m0}^q}{1-\Omega_{m0}^q}\right)\left( \frac{1-\tau}{n-\tau} \right)\,.
\end{equation}
Here the function $F_{vmpc}(z) = (\rho_m/\rho_{m0})^{q(n-1)}$ satisfies
\begin{equation}
    F_{vmpc} = (1+z)^{3 q(1-\tau)(n-1) }
    \bigg[\frac{\tilde\tau_2+1 }{\tilde\tau_2+F_{vmpc}}\bigg]^{\frac{\tau(n-1)}{\tau-n}} \,,
\end{equation}
and when $\tau = 0$, the Hubble parameter (\ref{vmpc:hubble}) becomes
\begin{equation}\label{fmpc:hubble}
    E^2 = \Omega_{m0}(1+z)^{3 }
      \left[1+ (\Omega_{m0}^{-q}-1)(1+z)^{3 q(n-1) }\right]^{\frac{1}{q}} \,.
\end{equation}

For the VEC model, $f = \exp{(e^{-ny})}$, and the solution is
\begin{equation}\label{VEC sol}
   \rho_m =  \rho_{m0}(1+z)^3 \left[\frac{n-(\rho_{m0}/\rho_{card})^{n } (1-\tau )}{n-(\rho_{m}/\rho_{card})^{n } (1-\tau )}\right]^\frac{\tau}{n (1-\tau )}\,,
\end{equation}
and the Hubble parameter is given by
\begin{equation}\label{vmpc:hubble}
    E^2 = \Omega_{m0}(1+z)^{3 }
    \bigg[\frac{\tilde\tau_3+1}{\tilde\tau_3+F_{vec}(z)}\bigg]^{\frac{\tau}{n(1-\tau)}}
    \exp{\bigg(-F_{vec}^{-1}~\ln\Omega_{m0}\bigg)} \,,
\end{equation}
where
\begin{equation}
    \tilde\tau_3 = \left( \frac{n}{1-\tau} \right)\ln\Omega_{m0}\,.
\end{equation}
Here the function $F_{vec}(z) = (\rho_m/\rho_{m0})^{n}$ satisfies
\begin{equation}
    F_{vec} = (1+z)^{3 n }
    \bigg[\frac{\tilde\tau_3+1 }{\tilde\tau_3+F_{vec}}\bigg]^{\frac{\tau}{1-\tau}} \,,
\end{equation}
and when $\tau = 0$, the Hubble parameter (\ref{vmpc:hubble}) becomes
\begin{equation}\label{fec:hubble}
    E^2 = \Omega_{m0}(1+z)^{3 }
        \exp{\bigg(-(1+z)^{-3 n }~\ln\Omega_{m0}\bigg)} \,.
\end{equation}
We summarize all the solutions of Hubble parameters for each model in Table \ref{table:models}.

\section{Statistical analysis with the observational data}

In general, the expansion history of the universe $H(z)$ or $E(z)$ can be given by a specific cosmological model or by
assuming an arbitrary ansatz, which may be not physically motivated but just designed to give a good fit to the data
for the luminosity distance $d_L$ or the 'Hubble-constant free' luminosity distance $D_L$ defined by
\begin{equation}\label{lu dis}
    D_L = \frac{H_0 d_L}{c} \,,
\end{equation}
where the light speed $c$ is recovered to show that $D_L$ is dimensionless. In the following, we will take the first
strategy that assuming the Hubble parameter $H(z; a_1, \cdots, a_n)$ with some parameters ($a_1,\cdots, a_n$) predicted by
the class of Cardassian models could be used to describe the universe, and then we obtain the predicted value of $D^{th}_L$ by
\begin{equation}\label{Dlth}
    D^{th}_L = \frac{(1+z)}{\sqrt{|\Omega_{k0}|}}\text{Sinn}\bigg[\sqrt{|\Omega_{k0}|}\int^z_0  \frac{dz'}{E(z';a_1, \cdots,
    a_n)}\bigg]
    \,,
\end{equation}
where $\Omega_{k0} = -k/(a_0^2H_0^2)$ and  $Sinn(x) = \sin(x), x, \sinh(x)$ for respectively a spatially closed
($\Omega_k<0$), flat ($\Omega_k=0$) and open ($\Omega_k>0$) universe.

On the other hand, the apparent magnitude of the supernova is related to the corresponding luminosity distance by
\begin{equation}\label{app mag}
   \mu(z) = m(z) - M = 5\log_{10}\left[\frac{d_L(z)}{\text{Mpc}}\right] + 25 \,,
\end{equation}
where $\mu(z)$ is the distance modulus and  $M$ is the absolute magnitude which is assumed to be constant for standard
candles like Type Ia supernovae. One can also rewrite the distance modulus in terms of $D_L$ as
\begin{equation}\label{app mag2}
    \mu(z) = 5\log_{10} D_L(z) + \mu_0 \,,
\end{equation}
where
\begin{equation}\label{zero off}
   \mu_0 = 5\log_{10} \left(\frac{cH_0^{-1}}{\text{Mpc}}\right) + 25 = -5\log_{10} h + 42.38 \,,
\end{equation}
is the zero point offset, which is an additional model independent parameter. Thus, we obtain the predicted value of
$\mu^{th}$ by using the value of $D_L^{th}$ and the observational value of $\mu^{obs}$ we used is the latest data
called the constitution data \cite{sne}, which contains $397$ data points including the the $307$ Union data set \cite{sneu} and $90$ CFA data set.

There are also some constraints from CMB and BAO observations. We will take the parameter $R$  from the CMB data \cite{cmb} and the parameter $A$  from the SDSS data \cite{sdss} as well as the supernova data to constrain parameters of the Cardassian models. The parameter $R$ is defined as
\begin{equation}\label{R def}
   R = \sqrt{ \frac{\Omega_{m0}}{|\Omega_{k0}|} } \text{Sinn} \bigg(\sqrt{|\Omega_{k0}|}\int^{z_{ls}}_0  \frac{dz'}{E(z')}\bigg) \,,
\end{equation}
where $z_{ls} = 1090$ is the redshift of the last scattering surface and the observational value is given by $R_{obs} = 1.170\pm0.019$. While the parameter $A$ is defined as
\begin{equation}\label{A def}
   A = \frac{\sqrt{ \Omega_{m0}} }{z_1} \bigg[ \frac{z_1}{ E(z_1)} \frac{1}{|\Omega_{k0}|}  \text{Sinn}^2 \bigg(\sqrt{|\Omega_{k0}|}\int^{z_{1}}_0  \frac{dz'}{E(z')}\bigg) \bigg]^{1/3} \,,
\end{equation}
where $z_1 = 0.35$ and the observational value is given by $A_{obs} = 0.469(0.95/0.98)^{-0.35}\pm 0.017$.

In order to determine the best value of parameters (with $1\sigma$  error at least ) in the Cardassian models, we
will use the maximum likelihood method and need to minimize the following quantity
\begin{equation}\label{tot chi2}
    \chi^2 = \tilde\chi^2_{SN} +\chi^2_{CMB} + \chi^2_{BAO} \,,
\end{equation}
where
\begin{equation}
    \chi^2_{CMB} = \bigg(\frac{R-1.710}{0.019}\bigg)^2\,, \quad \chi^2_{BAO} = \bigg(\frac{A-0.469(0.96/0.98)^{-0.35}}{0.017^2}\bigg)^2 \,,
\end{equation}
and
\begin{equation}\label{chi2}
    \tilde\chi^2_{SN}(a_1,\cdots,a_n) = \sum^{397}_{i=1} \frac{\bigg[\mu^{obs}(z_i) - 5\log_{10}D_L^{th}(z_i;a_1,\cdots,a_n
    ) - \mu_0 \bigg]^2}{\sigma_i^2} \,,
\end{equation}
where $\sigma_i$ is the $1\sigma$ error of the observation value $\mu^{obs}(z_i)$. Since the nuisance parameter $\mu_0$
is model-independent, then we analytically marginalize it by using a flat prior $P(\mu_0) = 1$:
\begin{equation}
    \chi^2_{SN} = -2\ln\left(\int_{-\infty}^{+\infty} e^{-\chi^2/2} P(\mu_0)d\mu_0\right) = a - \frac{b^2}{c} +
    \ln\left(\frac{c}{2\pi}\right) \,,
\end{equation}
where
\begin{equation}
    a = \sum^{397}_{i=1} \frac{\bigg[\mu^{obs}(z_i) - 5\log_{10}D_L^{th}(z_i;a_1,\cdots,a_n
    ) \bigg]^2}{\sigma_i^2} \,,
\end{equation}
\begin{equation}
    b = \sum^{397}_{i=1} \frac{\bigg[\mu^{obs}(z_i) - 5\log_{10}D_L^{th}(z_i;a_1,\cdots,a_n
    ) \bigg]}{\sigma_i^2} \,,
\end{equation}
and
\begin{equation}
    c = \sum^{397}_{i=1} \frac{1}{\sigma_i^2} \,,
\end{equation}
then, from now on, we will work with $\chi^2_{SN}$ and to minimize $\chi^2$ in eq.(\ref{tot chi2}). The best fit parameter values and the corresponding
$\chi^2_{\min}$ and $\chi^2_{min}/DOF$ will be summarized for each model in Table \ref{table:best}. Here DOF is the degree
of freedom defined as
\begin{equation}\label{dof}
    DOF = N - \nu \,,
\end{equation}
where $N$ is the number of data points, and $\nu$ is the number of free parameters.

\begin{figure}[h]
\begin{center}
\includegraphics[width=0.6\textwidth,angle=270]{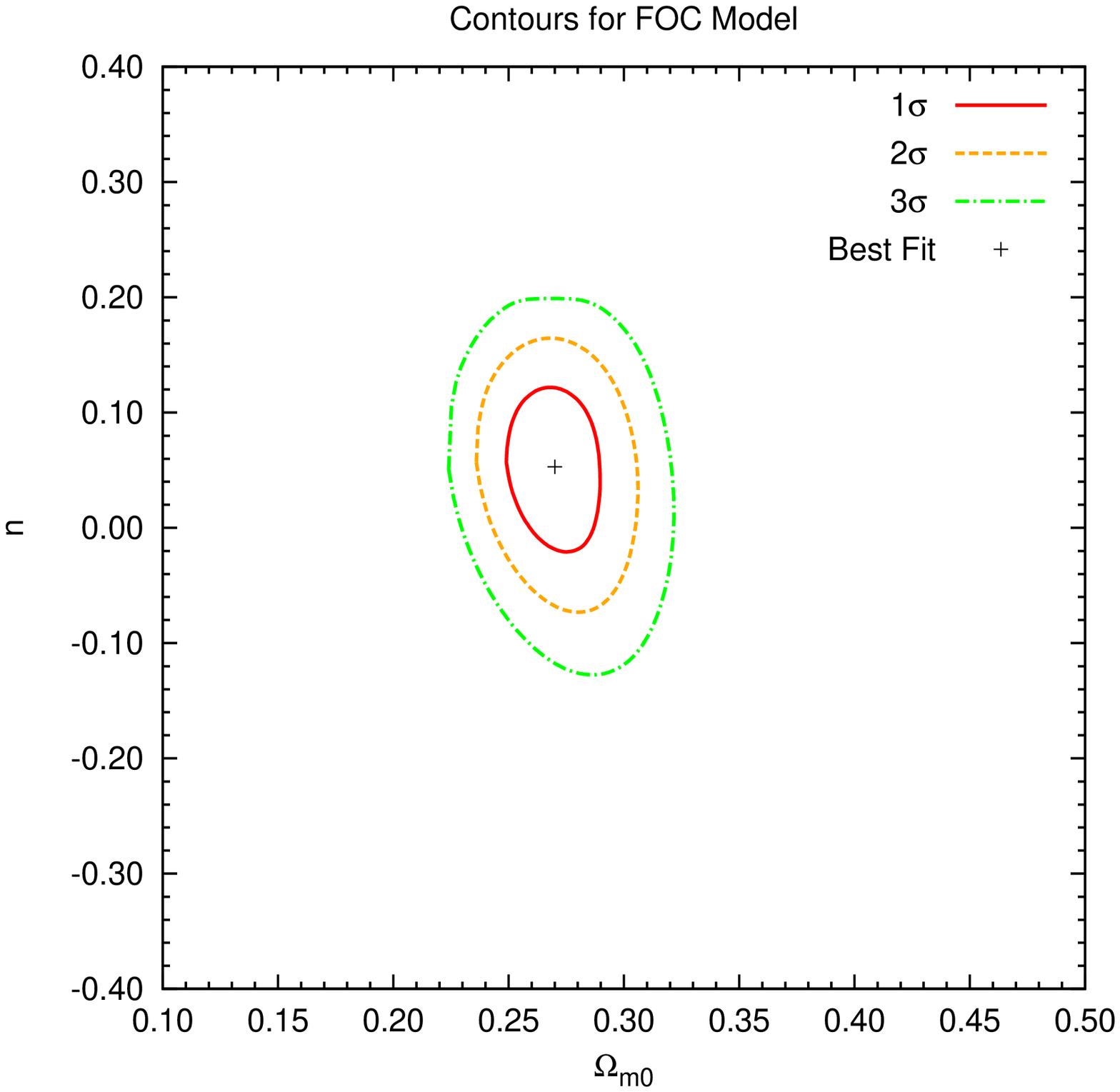}
\caption{\label{fig::foc}\textit{FOC:} Constraints on $\Omega_{m0}$ and $n$ from $1\sigma$ to $3\sigma$ confidence level obtained by using 397 SNe Ia + CMB + BAO for the FOC model. The best fit point corresponds to $\Omega_{m0} = 0.270$, $n = 0.053$.}
\end{center}
\end{figure}

\begin{figure}[h]
\begin{center}
\includegraphics[width=0.6\textwidth,angle=270]{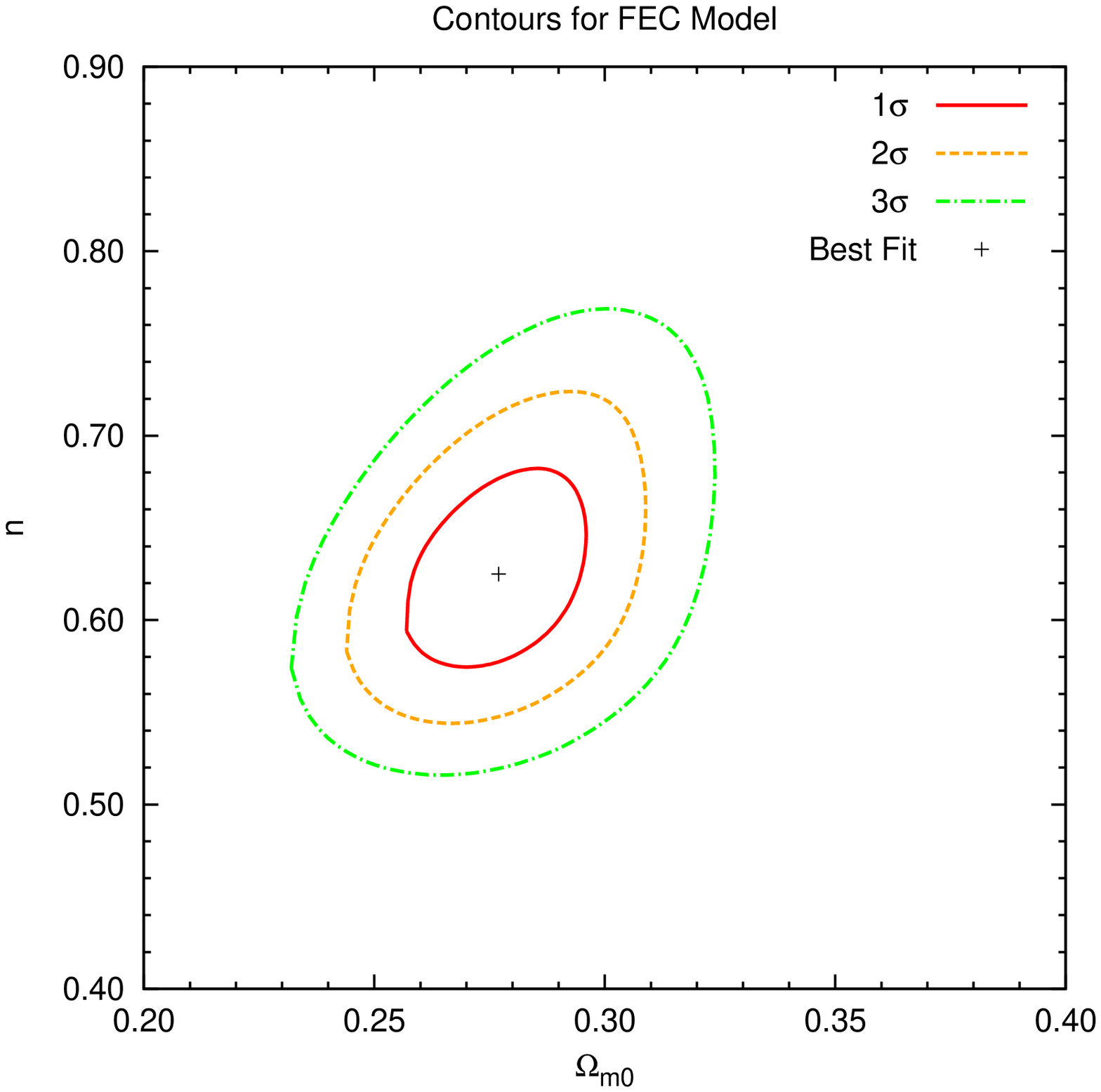}
\caption{\label{fig::fec}\textit{FEC:} Constraints on $\Omega_{m0}$ and $n$ from $1\sigma$ to $3\sigma$ confidence level obtained by using 397 SNe Ia + CMB + BAO for the FEC model.The best fit point corresponds to $\Omega_{m0} = 0.277$, $n = 0.625$.}
\end{center}
\end{figure}

\begin{figure}[h]
\begin{center}
\includegraphics[width=0.6\textwidth,angle=270]{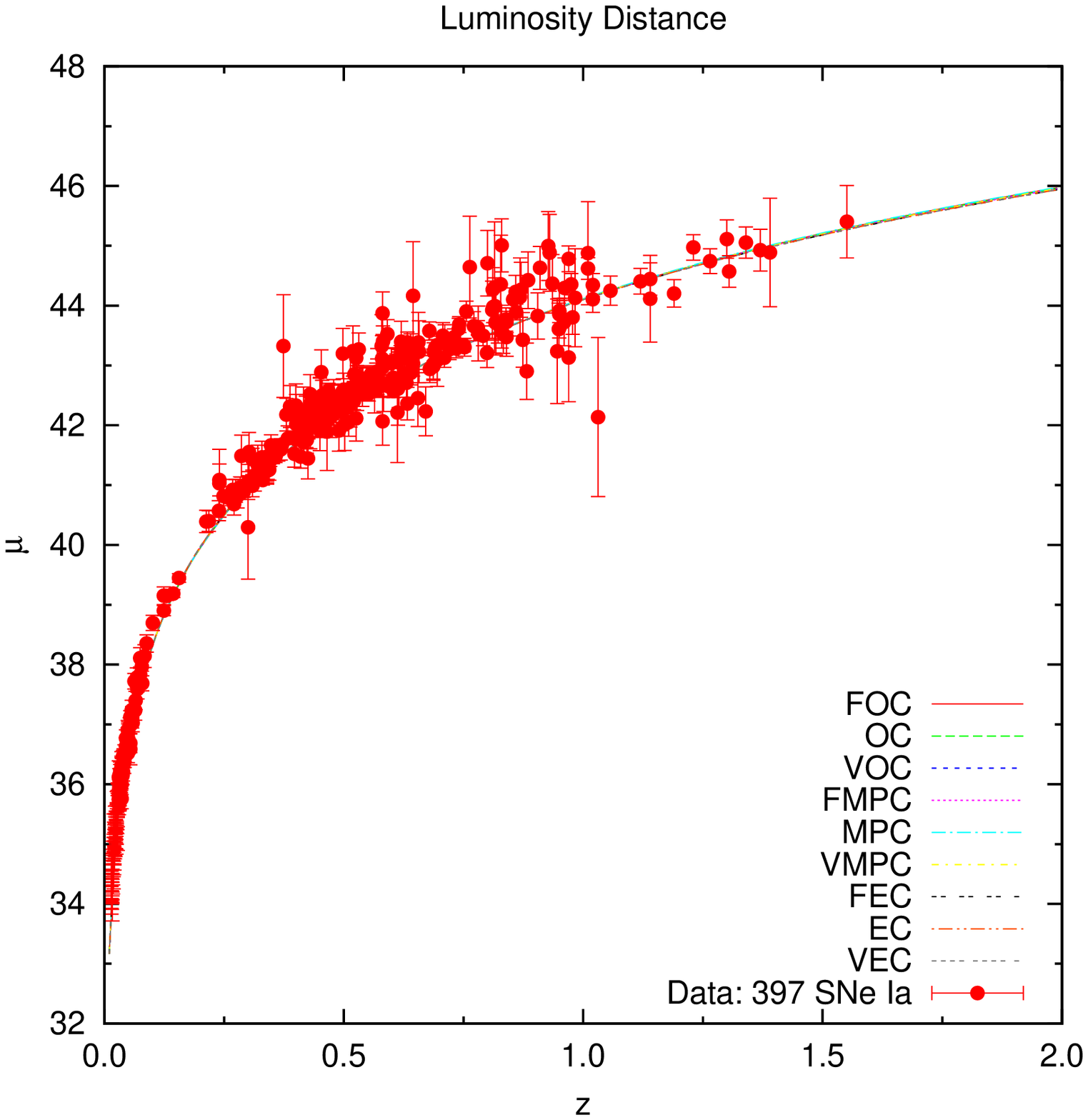}
\caption{\label{fig::dl}   The observed 397 SNeIa distance modulus along with the theoreti-
cally predicted curves in the Cardassan models with best fit parameters, and we have taken a \textit{priori} that current
dimensionless Hubble parameter $h = 0.70$.}
\end{center}
\end{figure}

We now apply the maximum likelihood method for each model in Table \ref{table:models}, and we summary the results in Table \ref{table:best} including the minimum values of $\chi^2$ and the best fit parameters with $1\sigma$ confidence level for each model. Since both the FOC and FEC model contain two parameters, we also plot the contours from $1\sigma$ to $3\sigma$ confidence levels for them, see Fig. \ref{fig::foc} and Fig. \ref{fig::fec}. For each model, the predicted dimensionless luminosity is plotted in Fig. \ref{fig::dl}, from which one can see that these Cardassian models  predict almost the same luminosity distance with their best fit parameters.

\begin{table}[h]
\centering
  \begin{tabular}{c|c|c|l}
  \hline
  \hline
  \multicolumn{1}{c|}{Model} & $\chi^2_{min}$ & $\chi^2_{min}/DOF$& \multicolumn{1}{c}{Best Fit Parameters ($1\sigma$)} \\
  \hline
  \hline
  FOC  & 474.083 & 1.194 & $\Omega_{m0} = 0.270^{+0.023}_{-0.021} $, $n = 0.053^{+0.070}_{-0.075} $                               \\
  \hline
  OC   & 473.084 &1.195  & $\Omega_{m0} = 0.283^{+0.037}_{-0.033} $, $n = 0.023^{+0.098}_{-0.113}$, $\Omega_{k0} = -0.010^{+0.018}_{-0.019}$             \\
  \hline
  VOC  & 473.101 & 1.195 & $\Omega_{m0} = 0.281^{+0.029}_{-0.031} $, $n = 0.010^{+0.110}_{-0.150} $,$\tau = -0.004^{+0.110}_{-0.150} $          \\
  \hline
  FMPC & 473.746 & 1.196 & $\Omega_{m0} = 0.273^{+0.027}_{-0.023} $, $n = -0.600^{+0.980}_{-0.450} $, $q = 0.480^{+2.020}_{-0.080}$             \\
  \hline
  MPC  & 473.072  & 1.198  & $\Omega_{m0} = 0.285^{+0.030}_{-0.035} $, $n = 0.200^{+0.200}_{-3.100}$, $q = 1.48^{+1.420}_{-1.280}$,  $\Omega_{k0} = -0.015^{+0.030}_{-0.015} $   \\
  \hline
  VMPC & 473.205 & 1.198 & $\Omega_{m0} = 0.279^{+0.026}_{-0.029} $, $n =-0.050^{+0.400}_{-2.950} $,   $q =0.900^{+2.000}_{-.0700} $, $\tau = -0.003^{+0.008}_{-0.007} $ \\
  \hline
  FEC  & 474.128  & 1.194 & $\Omega_{m0} = 0.277^{+0.024}_{-0.020}$, $n =0.625^{+0.059}_{-0.051} $                                   \\
  \hline
  EC   & 474.127 &1.197  & $\Omega_{m0} = 0.277^{+0.033}_{-0.027} $,  $n =0.626^{+0.094}_{-0.076} $ , $\Omega_{k0} =-0.0004^{+0.0213}_{-0.0196} $             \\
  \hline
  VEC  & 474.127 & 1.197 & $\Omega_{m0} = 0.276^{+0.034}_{-0.026}$, $n =0.623^{+0.117}_{-0.093} $,             $\tau = 0.0002^{+0.009}_{-0.0082}$           \\
  \hline
  \end{tabular}
  \caption{\label{table:best} Result: The minimum value of $\chi^2$ and the best fit parameters with $1\sigma$ confidence level in each model. }
\end{table}

\section{Discussion}
We have used the 397 SNe Ia, CMB and SDSS data to constrain several Cardassian models. We have summarized these model in Table \ref{table:models}, in which different forms of the function $g(\rho_m)$ have been chosen and the corresponding Hubble parameters are also given. In particular, we discuss the viscous Cardassian models in Section 2., in which we rewrite the Hubble parameter in a continent way to do the statistical analysis.

The fitting results are presented in Table \ref{table:best}, in which we have shown the minimum value of $\chi^2$ and the minimum value $\chi^2_{min}$ per degree of freedom.  The best fit parameters with $1\sigma$ confidence level for each model are also presented in Table \ref{table:best}, from which one can see that,  the latest observational data can not distinguish these models at this classical level. In other words, they predict almost the same evolution history of the universe and we need to take the perturbation of universe into account that will be studied in our further work.

In fact, the minimal of $\chi^2$ in eq.(\ref{tot chi2}) is very sensitive to the observational error of the distance modulus. Once the error is smaller in the future data than that at present, not every model will fit the data well, then one can distinguish these models and even rule out some of them. Thus, more precise data are very needed.

Since in Cardassian universe, one can expalain the accelerating expansion without introducing any dark energy component, it is very interesting and worth further studying. We also hope that future observation data could give more stringent constraints on the parameters in the Cardassian model.

\acknowledgments We thank Dao-Jun Liu and Ping Xi for useful discussions on the analysis of the data. This work is supported by National Science Foundation of China grant No. 10847153 and No. 10671128.

\end{document}